# Purcell effect of nitrogen-vacancy centers in nanodiamond coupled to propagating and localized surface plasmons revealed by photon-correlation cathodoluminescence

Sotatsu Yanagimoto[1], Naoki Yamamoto[1], Takumi Sannomiya[1, 2, *], Keiichirou Akiba[1,3, †]

[1] Department of Materials Science and Engineering, School of Materials and Chemical Technology, Tokyo Institute of Technology, 4259 Nagatsuta, Midoriku, Yokohama, 226-8503 Japan.

[2] JST-PRESTO, 4259 Nagatsuta, Midoriku, Yokohama, 226-8503 Japan.

[3] Takasaki Advanced Radiation Research Institute, National Institutes for Quantum and Radiological Science and Technology, 1233 Watanuki, Takasaki, Gunma, 370-1292, Japan

* sannomiya.t.aa@m.titech.ac.jp

† akiba.keiichiro@qst.go.jp

**ABSTRACT**

We measured the second-order correlation function of the cathodoluminescence intensity and investigated the Purcell effect by comparing the lifetimes of quantum emitters with and without metal structure. The increase in the electromagnetic local density of state due to the coupling of a quantum emitter with a plasmonic structure causes a shortening of the emitter lifetime, which is called the Purcell effect. Since the plasmon-enhanced electric field is confined well below the wavelength of light, the quantum emitter lifetime is changed in the nanoscale range. In this study, we combined cathodoluminescence in scanning (transmission) electron microscopy with Hanbury Brown-Twiss interferometry to measure the Purcell effect with nanometer and nanosecond resolutions. We used nitrogen-vacancy centers contained in nanodiamonds as quantum emitters and compared their lifetime in different environments: on a thin $SiO_2$ membrane, on a thick flat silver film, and embedded in a silver film. The lifetime reductions of nitrogen-vacancy centers were clearly observed in the samples with silver. We evaluated the lifetime by analytical calculation and numerical simulations and revealed the Purcell effects of emitters coupled to propagating and localized surface plasmons. This is the first experimental result showing the Purcell effect due to the coupling between nitrogen-vacancy centers in nanodiamonds and surface plasmon polaritons with nanometer resolution.

# I. INTRODUCTION

To implement quantum communications and optical computing, the transition probability of a quantum emitter (QE) is an important factor determining the speed of the information processing [1-3]. When a QE is coupled to an optical resonator or plasmonic structure, its transition probability can increase. This is known as the Purcell effect and is expected to be a promising way to speed up optical signal processing [4]. Since the coupling between the QE and the nanostructure is sensitive to the nanoscopic geometries, a visualization method of photonic information with nanoscale resolution is required. The Purcell effect can be evaluated by using changes in the lifetime of the QE, which is usually measured with a laser pulse excitation and synchronized detection. Among various proposed QEs, nitrogen-vacancy (NV) emission centers in nanodiamonds (NDs) are considered a promising candidate due to their high stability and various functions [5,6]. However, the lifetime of ND emitters is strongly dependent on their dielectric environment, i.e., the ND structure itself [7-9], and therefore, it is necessary to measure individual NDs to evaluate the lifetime. To measure the lifetime of individual NDs by fully photon-based methods, they must be separated more than the diffraction limit of light and nanoscopic measurement of a sub-wavelength object is not straightforward. Although super-resolution microscopy, such as stochastic optical reconstruction microscopy (STORM) [10] and stimulated emission depletion (STED) microscopy [11], is becoming more popular, the stochastic approach is not suited for lifetime measurement because it requires spatially resolved detection and the depletion approach could possibly modify the lifetime of the target object due to the scattering of the depletion light by nanostructures.

Besides purely optical means, cathodoluminescence (CL) microscopy, in which the nano-resolution electron beam of a scanning (transmission) electron microscope (S(T)EM) is utilized for the excitation of light emission sources, has become popularly used as a powerful tool to measure nanoscale optical characteristics [12,13]. This method enables evaluating the optical properties of the materials and nanostructures well beyond the diffraction limit of light. For instance, localized surface plasmon (LSP) and surface plasmon polariton (SPP) have been visualized [8,14,15]. The lifetime measurement using CL typically requires a large-scale measurement system that integrates a pulsed electron gun as an excitation source [16,17] and a synchronized detector, which is especially complicated for STEM with high acceleration voltages.

As an alternative to using a pulsed electron gun, the lifetime measurement based on the Hanbury Brown-Twiss (HBT) interferometer combined with the STEM/SEM-CL has been recently proposed [18-21]. Since HBT-based measurement is a fully passive method that requires modifications only in the light detection system without pulsing the electron source, such a setup can be readily implemented to higher voltage microscopes, like STEM. This type of method has been applied to QEs at low temperatures [18,21] and quantum wells at room temperature [19,22]. The Purcell effect of the coupling between QEs and LSPs has also been evaluated at low temperature [21]. However, no Purcell effect based on the coupling to propagating SPPs has been experimentally observed in the previous work using NV centers in NDs as QEs [20], while such coupling between a dipole and SPPs has been theoretically predicted [23]. In a practical device application, SPPs are actually attractive since they can both introduce the Purcell effect and transfer the information as a propagating surface wave. This should provide a great potential to realize on-chip devices with QEs coupled to SPPs communicating with other optical elements in a distance. Therefore, it is important to confirm if the coupling of a QE to an SPP is possible as in the theory.

In this paper, we experimentally clarify this controversial coupling of QEs and SPPs by measuring individual QEs using the HBT-based STEM-CL. The lifetime is obtained through photon bunching in the correlation measurement, which can be analytically described with a simple two-level system. We used NV centers in NDs as a typical candidate of QEs. As the simplest coupling system, we firstly analyzed NVs coupled to SPPs on a flat metal surface, and then in addition, we also propose a more efficient coupling of QEs to SPPs and LSPs by embedding NDs in metal films. The Purcell effect based on the coupling between QEs and SPPs or LSPs was experimentally proved with clear shifts in the lifetime histograms of individual NDs. We evaluated the Purcell effect in detail with both analytical and numerical methods providing insights into the coupling dependence on the wavelength, orientation, and location of the emitter dipole.

## II. METHOD

### A. Sample fabrication

ND solution (900174-5ML, Sigma-Aldrich, America) containing 900 or more NV centers was used. The average diameter of the NDs is 100 nm. Three types of samples (Samples A, B, and C) as illustrated in Fig. 1 are prepared. NDs were dispersed on a 30 nm thick free-standing $SiO_2$ membrane for Sample A and on a 300 nm thick silver film thermally deposited on an InP substrate for Sample B. For Sample C, NDs were dispersed

on a 30 nm SiO$_2$ thin membrane, followed by sputter-deposition of a 300 nm Ag film to embed the NDs in the

Ag film. More fabrication details are shown in the Supplemental Material [24]. In all the measurement, the electron beam is incident and the optical signal is detected from the upper side in Fig 1. Sample A without metal was used as a reference to evaluate the lifetime change in Samples B and C.

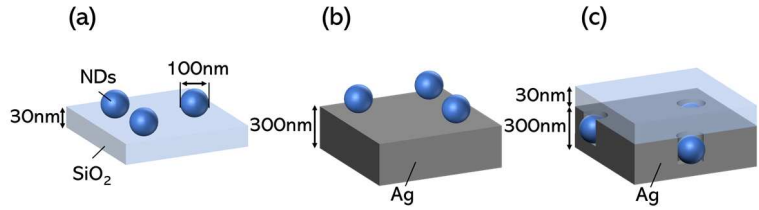

FIG.1. Schematics of the fabricated samples. The samples were prepared by drop-casting

the ND solution onto (a: Sample A) a 30 nm thick SiO$_2$ membrane and (b: Sample B) a

300 nm thick Ag film on an InP substrate. (c) Sample C is prepared by depositing a 300

nm thick Ag film on a 30 nm SiO$_2$ membrane with ND particles on it. All samples were placed on the Cu grid.

**B. Lifetime measurement by CL**

The lifetime of the QE $\tau_0[s]$ is the reciprocal of the total decay rate $\gamma_0[s^{-1}]$. The total decay rate can be expressed as a sum of the decay rates due to photon emission ($\gamma_r$), absorption into the surrounding environment

($\gamma_n$), and plasmon excitation ($\gamma_p$),

$$\gamma_0 = \gamma_r + \gamma_n + \gamma_p. \tag{1}$$

The decay due to plasmon excitation is proportional to the electromagnetic local density of state (EMLDOS) near the QE [1,23,26]. EMLDOS is the number of electromagnetic modes per unit volume coupled with the harmonic oscillating mode of the QE, which increases with the addition of plasmon mode, for example, if there is a metal surface near the QE. The Purcell effect is generally quantified and evaluated by the Purcell factor $F_p = \gamma_0/\gamma_0^{free}$, where $\gamma_0^{free}$ is the decay rate in free space. The lifetime $\tau_0$ is obtained by measuring the second-order correlation function $g^{(2)}(\tau)$ using an HBT interferometer (Fig. 2 (a)). $g^{(2)}(\tau)$ is defined by

$$g^{(2)}(\tau) = \frac{\langle I(t)I(t+\tau)\rangle}{\langle I(t)\rangle\langle I(t+\tau)\rangle},\tag{2}$$

where $I(t)$ is the CL signal intensity as a function of time $t$ and $\tau$ denotes the time delay. The HBT interferometer consists of a system that divides the CL signal into two paths and measures the $g^{(2)}(\tau)$ of the incident photons using single-photon counting modules (SPCMs). We used a CL detection system installed in a STEM (JEM-2000FX, JEOL, Japan) equipped with a thermal emission gun to measure the lifetime of individual NDs, as shown in Fig. 2. The light emitted from the sample is guided out of the STEM column by a parabolic mirror, and transferred either to a spectrometer for CL spectrum measurement, to a photo-multiplier tube (PMT) for panchromatic CL mapping, or to the HBT measurement system. The thermal photon radiation from the emitter is avoided in CL detection by letting the direct reflection from the substrate going through the hole of the parabolic mirror for the electron beam path and by subtracting the stray signal as background. The background signal becomes a constant offset in the $g^{(2)}(\tau)$ measurement, which does not affect the lifetime measurement. We performed the measurement at room temperature at an accelerating voltage of 80 kV and a beam current of between 17 to 60 pA, which forms a probe of 10 nm or less with the convergence half angle of about 1mrad. We confirmed that this condition is feasible for the lifetime measurement of NDs to obtain clear bunching features in the $g^{(2)}(\tau)$ correlation curve within 200 seconds of exposure (see Supplemental Material for the current dependence [24]). The CL spectrum from a single ND of Sample A without a metallic structure is shown in the inset of Fig. 2. We observed the emission from the $NV^0$ center, which has a zero-phonon line (ZPL) at a wavelength of 575 nm [6], but without the charged center ($NV^-$). This result agrees with the previous report [27] that only the emission from $NV^0$ can be observed in CL. Thus the $g^{(2)}(\tau)$ measurement using CL in this paper provides information on the lifetime of $NV^0$ centers in NDs.

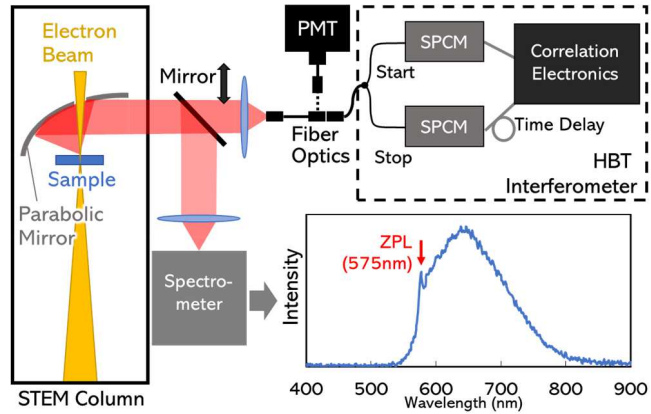

FIG.2. Schematic illustration of the CL-STEM and optical setup for the HBT measurement. A parabolic mirror collimates the emitted light that is transferred either to a spectrometer, photo-multiplier tube (PMT), or HBT interferometer, which is selected by the movable mirror and optical fiber connection. Inset: CL spectrum of an ND without the metal

structure (Sample A) measured at room temperature with the exposure time of 10 seconds.

The zero-phonon line (ZPL) is indicated by the red arrow.

## III. EXPERIMENTAL RESULTS

Figure 3 shows the SEM or STEM (bright field) images (left panel), CL-maps (center panel), and $g^{(2)}(\tau)$ curves (right panel) for representative ND particles in the same field of view for each sample. The selected ND particles are indicated by circles, of which color corresponds to the $g^{(2)}(\tau)$ curve. The experimental lifetime $\tau_0^{Exp}$ of each sample was obtained by fitting the correlation plot with $g^{(2)}(\tau) = 1 + A \exp(-|\tau|/\tau_0^{Exp})$, where the coefficient $A$ equals to $(g^{(2)}(0) - 1)$ [18,24]. This expression of the $g^{(2)}(\tau)$ function can be analytically derived based on the rate equation of a two-level system (see Supplemental Material [24]). The photon bunching occurs even at room temperature, however, with shorter lifetimes compared to those under liquid $N_2$ cooling condition, which is about 20 ns on average [18].

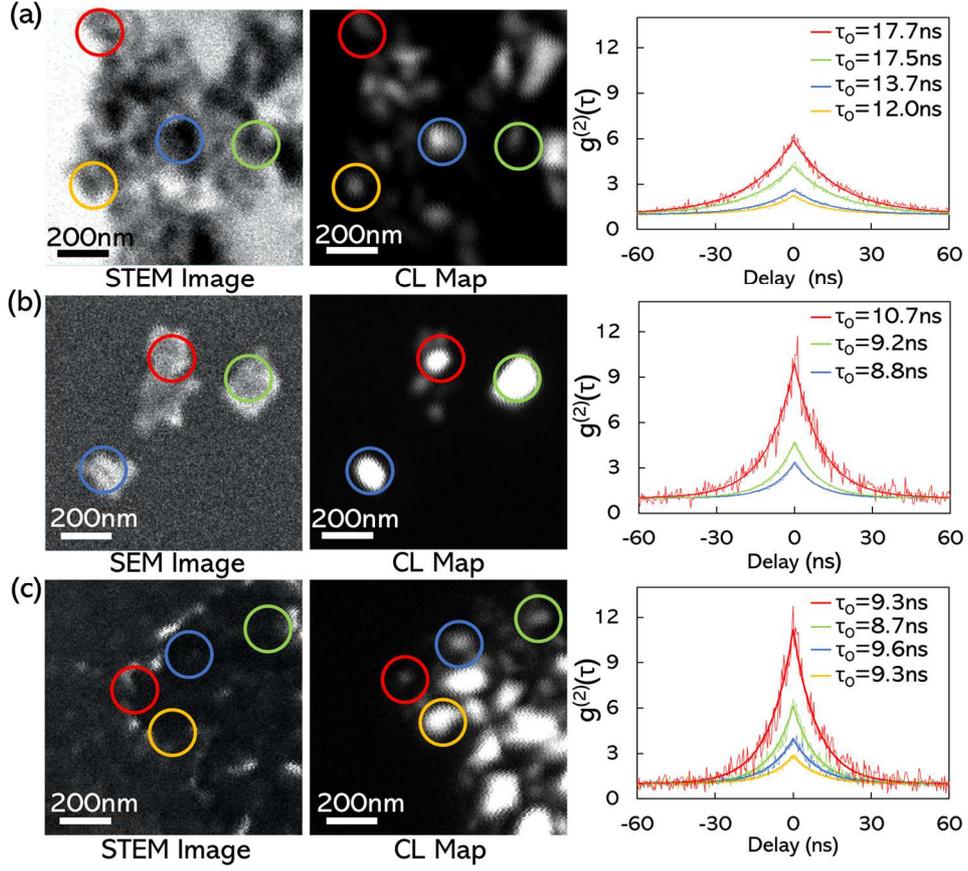

FIG.3. CL-HBT measurement results of Samples (a) A, (b) B, and (c) C, respectively.

From left to right: SEM or STEM image, CL panchromatic image, and $g^{(2)}(\tau)$ plots with fitted curves for NDs in each sample. The color of $g^{(2)}(\tau)$ curve in each graph corresponds to the color of the circle in the map indicating the measured NDs.

The value of $(g^{(2)}(0) - 1)$ is inversely proportional to the beam current of the incident electron [19]. This relation between $(g^{(2)}(0) - 1)$ and the beam current deduces the excitation probability, which is shown in

the Supplemental Material [24]. Even without metal structure, the values of $g^{(2)}(0)$ and lifetime of bare

NDs in Sample A vary significantly from particle to particle. This is due to differences in the shape of the particles, distribution of NV centers within the particles, and the surrounding environment [28,29]. We investigated the size dependence of the Purcell factor using electromagnetic field simulation and revealed that the Purcell factor can ~~change~~ vary ~~with~~by the particle size. The details of this size effect simulation are discussed in the Supplemental Materials [24]. We also observed individual emission spectra from several NDs and found more spectral variations for Samples B and C with metal structures compared to bare ones in Sample A (see Fig. S7 in the Supplemental Material [24]). In spite of the wide distribution of the lifetime values, we find a trend that the lifetime is shorter and $g^{(2)}(0)$ is larger for Samples B and C compared to the bare NDs in Sample A. To reliably evaluate this tendency and distribution of the lifetime, we measured 53, 58, and 55 NDs for Samples A, B and C, respectively and subsequently made a histogram of the lifetime, as shown in Fig. 4. The lifetime distribution of individual NV particles is presented in the bottom part.

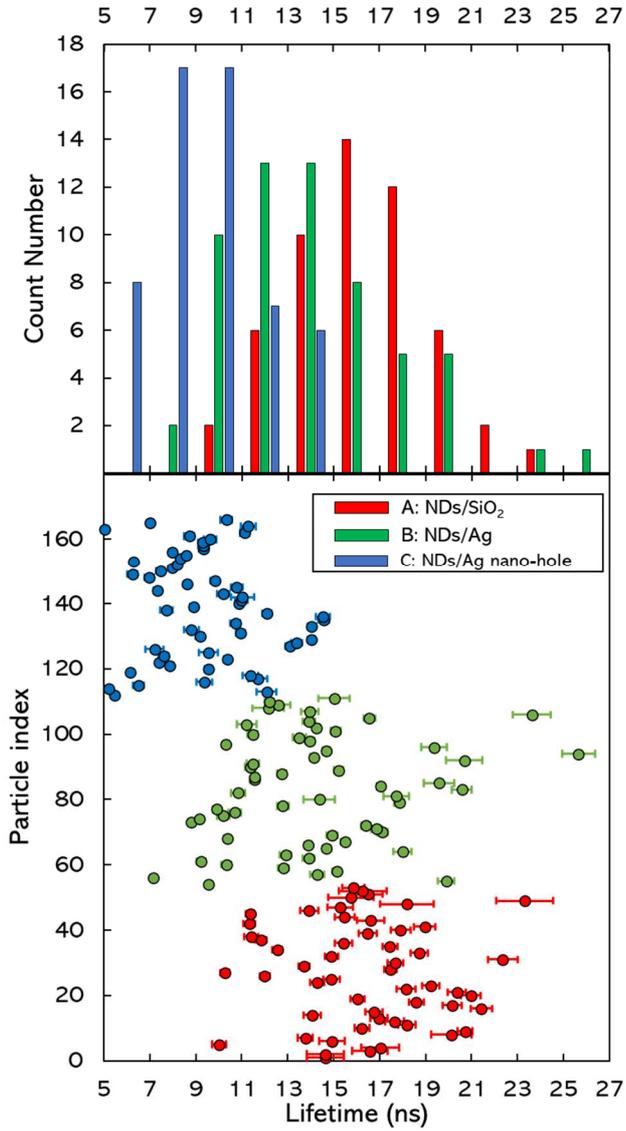

FIG.4. Histograms (top) and scatter plots (bottom) of the lifetime measurement of Sample

A (red), B (green), and C (blue). The bin width for the histogram is set to 2 ns. The error bars of individual lifetime measurement are shown in the scatter plot.

The histogram and scatter plot clearly show that the lifetime distributions with metallic structures (Samples

B and C) are shifted to shorter lifetime values compared to Sample A. The shift was particularly pronounced

in Sample C. To confirm the distribution difference statistically and quantitatively, we performed Wilcoxon-

Mann-Whitney statistical u-test [30]. This u-test is a method of determining the probability that two statistical data are extracted from the same population. If the probability is small enough, it is determined that the shift of the lifetime distribution is not due to statistical fluctuation, but in this case due to the Purcell effect. We run the test at the 5% level of significance and the probability between A-B, A-C and B-C are calculated to be $p_{AB}=4.39\times10^{-4}$, $p_{AC}=1.13\times10^{-16}$, and $p_{BC}=1.84\times10^{-11}$, respectively. All results are well below the significance, which means a substantial separation of the distributions. This result justifies our statistical evaluation of the Purcell effect, allowing us to compare the statistical value of each sample. Table.1 shows the Purcell factors calculated from the mean, mode, and median of each sample data. All the statistical values gave the same tendency of the lifetime, and therefore, we used the Purcell factor calculated from the mean for the following discussion. The mean values of the Purcell factor of Sample B and C are $F_p^{ExpB}=1.15$, and $F_p^{ExpC}=1.73$, respectively. These values again verify that the NV centers embedded in metal (Sample C) are more affected by the Purcell effect than those on a flat metal substrate (Sample B).

TABLE I. Lifetime and Purcell factor of different statistical values

| Sample | Mean | | Mode | | Median | |
|---|---|---|---|---|---|---|
| | Lifetime (ns) | Purcell factor | Lifetime (ns) | Purcell factor | Lifetime (ns) | Purcell factor |
| A | 16.3 | | 16 | | 16.5 | |
| B | 14.2 | 1.15 | 13 | 1.23 | 13.9 | 1.19 |
| C | 9.4 | 1.73 | 9 | 1.78 | 9.3 | 1.77 |

We find that the lifetime distribution is narrower for Sample C with the shorter lifetime compared to Sample A. This tendency of lifetime distribution is related to an additional decay rate, i.e., $\gamma_p$, which can be confirmed by modeling a certain decay rate distribution and shifting it by a constant $\gamma_p$. The resultant lifetime distribution becomes narrower in the time domain (see Supplemental Material for details on the change of distribution [24]).

We also notice the tendency that the $g^{(2)}(0)$ values are larger in the samples with metal structures, which are represented by 4.09, 6.93, and 7.11 as mean values for samples A, B, and C under the same excitation condition, respectively. This can be attributed to the increase of $\gamma_0$ or decrease of $\tau_0$ because the $g^{(2)}(0)$ increases with $\gamma_0$ (see Eq.(S15) in the Supplemental Material [24]). In addition, the $g^{(2)}(0)$ value may also involve the effect of the secondary electron emitted from supporting materials by electron irradiation: More secondary electrons are expected to be emitted in Samples B and C than Sample A since the bulk Ag substrate can produce more secondary electrons than the thin free-standing $SiO_2$ membrane in Sample A. The increase of the luminescence intensity in samples B and C can be due to the excitation of the secondary electrons. Therefore, the secondary electrons from the Ag substrate probably can excite a large number of NV centers and, as a result, may have enhanced the $g^{(2)}(0)$ values for samples B and C.

## IV. DISCUSSION

### A. Coupling of an electric dipole to a flat metal surface

We have experimentally shown the Purcell effect of emitters of NDs coupled to a flat metal surface (Sample B). This result is different from the previous report, where no lifetime change was observed for NDs on a flat Ag flake [20], probably because the lifetime distribution of their original NDs is too large, concealing the distribution shift. In this section, we analyze such coupling by considering the simplest model that consists of an electric dipole near a metal surface. We assume a system with two domains, i.e., dielectric (domain 1) and metallic (domain 2) half-spaces, and an electric dipole in the dielectric half-space. The definitions of the model and the axis coordinate are illustrated in the inset of Fig. 5. The media in both domains are homogeneous, isotropic, and linear. The refractive indices are set to $n_1 = \sqrt{\varepsilon_1}$ and $n_2 = \sqrt{\varepsilon_2}$ for domains 1 and 2 respectively, where $\varepsilon_i$ is the relative permittivity of domain $i$. Here, we set domain 1 as vacuum ($n_1 = 1$) or diamond ($n_1 = 2.42$). For the relative permittivity of domain 2, the literature value of Ag is used [31]. In this model, the Purcell factor of an electric dipole with a dipole moment $\boldsymbol{\mu}$ located at $z = z_0$ is expressed as

$$\frac{\gamma_0^{sub}}{n_1\gamma_0^{free}} = 1 + \frac{\mu_x^2+\mu_y^2}{|\boldsymbol{\mu}|^2}\frac{3}{4n_1^{\frac{3}{2}}}\int_0^{\infty} Re\left\{\frac{s}{s_z}\left[r_s - s_z^2 r_p\right]e^{2ik_1z_0s_z}\right\}ds + \frac{\mu_z^2}{|\boldsymbol{\mu}|^2}\frac{3}{2n_1^{\frac{3}{2}}}\int_0^{\infty} Re\left\{\frac{s^3}{s_z} r_p e^{2ik_1z_0s_z}\right\}ds, \qquad (3)$$

where the $z$-axis is perpendicular to the surface, and $r_s$ and $r_p$ are the Fresnel reflection coefficients for $s$- and $p$-polarizations, respectively [23]. For convenience, we introduced $s = k_\rho/k_1$ and $s_z = \sqrt{1-s^2} = k_{z1}/k_1$, where $k_\rho$ is the wavenumber component parallel to the interface. The dispersion relations of the SPP on the Ag surface is expressed as $s = \sqrt{\frac{\varepsilon_1\varepsilon_2}{\varepsilon_1+\varepsilon_2}}$ ($s = 1$ corresponds to the light line). Therefore, the decay rate due to radiation, absorption, and excitation of the SPP can be separately obtained by integrating Eq. (3) over appropriate ranges of $s$. Here, we used the integration ranges according to Weber and Eagen [32]. By using Eq. (3), the Purcell factors of dipoles in vacuum ($n_1 = 1$) set horizontally and vertically to the interface were calculated at the wavelength of 575 nm as a function of the vertical position of the dipole from the interface $z_0$, as shown in Fig. 5.

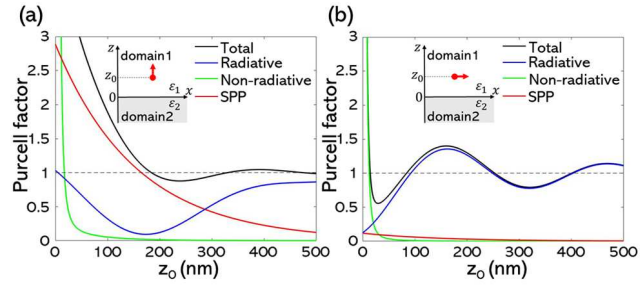

FIG.5. Analytically calculated Purcell factors of a dipole in vacuum on a flat Ag surface

as a function of the distance from the surface. (a) Vertical ($z$-) and (b) horizontal ($x$-)

dipoles placed on the Ag surface ($z$-dipole : $\mu_x = \mu_y = 0, \mu_z = 1$, $x$-dipole : $\mu_x = 1$,

$\mu_y = 0, \mu_z = 0$).

The total decay rate of the dipole is sensitive to position within a distance less than $z_0 = 100$ nm. We notice that the non-radiative decay rate due to absorption to the Ag substrate (green line) does not strongly depend on the direction of the dipole moment and exhibits a monotonous decrease with distance. In contrast, the interference between the radiative field from the dipole and the reflection field from the Ag surface causes oscillating features in the radiative decay rate, which is wavelength dependent. The total Purcell factor converges to 1 at the limit of the infinite separation distance $z_0 \to \infty$. The decay rate due to the SPP excitation

$\gamma_{SPP}$ of the vertical dipole is larger than that of the horizontal dipole at all distances because the vertical dipole

can more efficiently excite SPPs by inducing surface charges with the electric field perpendicular to the metal surface.

Photons detected in the HBT experiment have various wavelengths due to the broad spectrum, which must be considered when calculating the Purcell factor. We now assume that the dipoles give the same radiation spectrum as shown in Fig. 2(b) and are homogeneously distributed inside a spherical ND particle. On the basis of this model emulating the experiment, we can calculate the averaged Purcell factors. The average Purcell factor $F^{Int}$ is obtained by averaging the Purcell factors over the wavelength range of the spectrum and the volume (or $z$ position). This weighted average can be expressed as:

$$F^{Int} = \frac{6}{\pi D^3} \frac{\int_{z_0} \int_{\lambda} \pi z_0 (D - z_0) I_{Cl}(\lambda) \gamma_{rad}^{sub}(\lambda) \gamma_0^{sub}(\lambda, z_0) / \gamma_0^{free}(\lambda) d\lambda \, dz_0}{\int_{\lambda} I_{CL}(\lambda) \gamma_{rad}^{sub}(\lambda) \, d\lambda} \tag{4}$$

where $I_{Cl}(\lambda)$ is the CL intensity shown in Fig. 2(b) and is linear to the number of emitted photons per unit time. We used the diameter of the ND particle $D = 100 [nm]$. Since the Purcell factor diverges when $z_0 = 0$ nm, as seen in Fig. 5, we assume a gap of 5 nm between the particle and the substrate. This thickness is roughly corresponding to the sum of the typical oxide layer thickness on the Ag surface and the amorphous (or graphitized) layer thickness in ND [24,33,34]. Table 2 shows the calculated results of the Purcell factors averaged over both wavelength and volume when the dielectric domain is diamond ($n_1 = 2.42$) and vacuum ($n_1 = 1$). In both cases, the total Purcell factor exceeds 1, showing that the decay rate is enhanced by the presence of the Ag substrate. This analytical evaluation gives ideas of the Purcell effect of a flat metal surface such as the case of Sample B although we here do not correctly consider the presence of a dielectric sphere around the dipole in the ND particle.

TABLE II. Purcell factors of a dipole on a metal surface calculated with the analytical formula

| Domain1 | Purcell factor | | |
| --- | --- | --- | --- |
| | Horizontal ($x$-) dipole | Vertical ($z$-) dipole | Total |
| Diamond | 1.12 | 1.35 | 1.20 |
| Vacuum | 0.66 | 2.88 | 1.56 |

## B. FEM simulations

In this section, we will numerically process the dipoles in finite-sized dielectric spheres in a more complex environment such as Sample C, where the analytical approach is no longer appropriate. We performed electromagnetic field simulations with a finite element method (FEM) using COMSOL Multiphysics and numerically evaluated the Purcell factors considering the dielectric effect of the sphere. FEM can also provide electric field distributions that give insights into EMLDOS and the resonance modes that couple to the emitter dipole [35-37].

Three models A, B, and C, as shown in Fig. 6, are prepared according to the experiment. The Purcell factors are calculated using Model A as a reference and compared with the experimental results. An electric point dipole with a dipole moment $|\boldsymbol{\mu}| = 1[C \cdot m]$ is placed in a spherical ND particle with a diameter of 100 nm, and the entire system is surrounded by a perfectly matched layer (PML). In all models, the ND particle is covered by a 2.5 nm thick graphitized layer, which was experimentally confirmed [24] and also previously reported [34]. The calculation cell size is set to double the wavelength of the radiation field, and the thickness of the PML was set to 60% of the wavelength [35]. Although we have seen the analytical results that the Purcell factor depends on the distance from the substrate as shown in Fig. 5, we here present only dipoles placed in the center of the particle as the representative case. The results of the dipoles at different positions are shown in the Supplemental Material [24]. The wavelength-dependent Purcell factors were calculated for the emitter dipole moments in the $z$-axis and $x$-axis directions. The refractive index of diamond was set to 2.4 and the literature values of the dielectric function of Ag [38], graphite [39], and SiO$_2$ [40] were used.

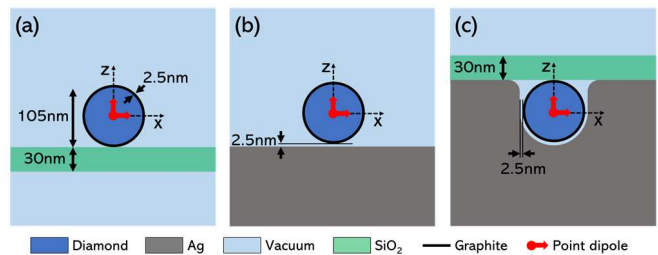

FIG.6. FEM models. A 100 nm diameter diamond sphere covered with a 2.5 nm thick graphite layer is placed (a) on a 30 nm thick SiO$_2$ membrane (Model A), (b) on a Ag substrate (Model B), and (c) in a Ag hole with a rounded edge with a radius of curvature of 25 nm [41]. For Model C, the Ag hole is covered with a 30 nm thick of SiO$_2$ membrane to emulate Sample A. An electric point dipole is placed at the center of the sphere. The results with the excitation electric dipoles in different positions are found in the Supplementary Material [24].

By considering that carbon has poor adhesion to Ag and that the particles used in the experiment are not perfectly spherical [24], it is expected that there is a gap between the Ag substrate and the particle. In addition, an Ag oxide layer covers the surface of the Ag substrate [24,33,34]. Therefore, for models B and C, a 2.5 nm gap between the particle and Ag surface was introduced. This gap helps avoid field singularities at the triple interface points. The Purcell factors were calculated from the FEM results in the following procedure [23,35]. The electric field can be described as the sum of the direct radiation field $\boldsymbol{E_0}$ from the excitation dipole and the scattering field $\boldsymbol{E_s}$ due to the surrounding structure:

$$\boldsymbol{E(r,\omega)} = \boldsymbol{E_0(r,\omega)} + \boldsymbol{E_s(r,\omega)} \tag{5}$$

The energy dissipation $P_0(\omega)$ from the excitation electric dipole with a finite volume $V$ located at $\boldsymbol{r} = \boldsymbol{r_0}$ at an angular frequency $\omega$ can be expressed by the following equation,

$$P_0(\omega) = -\frac{1}{2} \int_V \boldsymbol{Re}\{\boldsymbol{j^*(r_0)} \cdot \boldsymbol{E(r,\omega)}\} \, dV, \tag{6}$$

where $\boldsymbol{j(r_0)}$ is the current density induced by the excitation electric dipole at $\boldsymbol{r} = \boldsymbol{r_0}$ (* denotes the complex conjugate) and is expressed by

$$\boldsymbol{j(r_0)} = -i\omega\boldsymbol{\mu(r_0)} \, . \tag{7}$$

By using Eq. (6) and (7), we obtain the following equation,

$$P_0(\omega) = \frac{\omega}{2} \int_V \boldsymbol{Im}\{\boldsymbol{\mu^*(r_0)} \cdot \boldsymbol{E(r,\omega)}\} \, dV. \tag{8}$$

Since the decay rate is linear to the energy dissipation, we define the Purcell factors of models B and C in reference to model A : $F_0^{SimB,C} = \gamma_0^{SimB,C}(\omega)/\gamma_0^{SimA}(\omega) = P_0^{SimB,C}(\omega)/P_0^{SimA}(\omega)$.

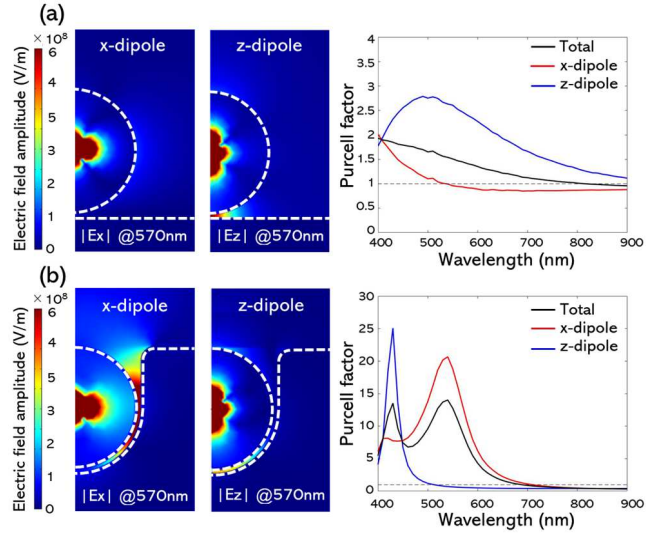

FIG.7. FEM simulation for (a) Model B and (b) Model C. From left to right: *x*-component electric field map with the *x*-dipole excitation, *z*-component electric field map with the *z*-dipole excitation, and the Purcell factor as a function of the wavelength, respectively. The excitation electric dipole is located at the center of the ND particle and the Purcell factor is calculated by integrating the inner product of the excitation dipole moment and the amplitude of the generated electric field. For the mapping, only half of the space is shown, based on the symmetry.

Figure 7 shows the calculation results of the electromagnetic field simulation with the excitation electric dipole located at the center of the ND particle. According to Eq. (8), the Purcell factor spectra are calculated by volume integration of the inner product of the excitation dipole moment and the amplitude of the generated electric field. For model B, the electric field is enhanced at the gap between the particle and the substrate, especially for the dipole orientation along the *z*-axis (*z*-dipole). Also, the Purcell factor is larger for the *z*- dipole than the *x*-dipole. We note that this result is consistent with the analytical calculation of a dipole on a flat metal surface as expressed in Eq. (3). Similar to the analytical calculation, the Purcell factors exhibit a strong dependence on the dipole orientation due to the coupling strength with the structure, i.e., the coupling to SPPs. The spectral response in Fig. 7(a) is indeed similar to the analytical one with a dipole placed 50 nm above the silver surface (see Supplemental Material [24]).

For model C, as shown in the field maps in Fig. 7(b), the electric field in ND is more enhanced compared to model B. This enhancement of EMLDOS is more remarkable with the *x*-oriented dipole. Indeed, the calculated Purcell factors are stronger for *x*-dipole excitation than for *z*-dipole, as shown in Fig. 7(b). We notice the LSP resonance peaks in the spectra: an in-plane mode at 540 nm with *x*-dipole excitation and an out-of-plane mode at 430 nm with *z*-dipole excitation. Since this model also consists of a continuous metal surface as Model B, the Purcell factor must be derived from both SPP and LSP. Since we were not able to experimentally capture the wavelength dependent Purcell factors due to the limited emission wavelength range of NV centers, we calculated the intensity-averaged Purcell factors similar to Eq. (4) with the exception of the volume average to evaluate the average response over the whole spectral range. The averaged results are shown in Table.3. Here, the radiative portion of the weighted average, as in Eq. (4), was obtained by integrating the Poynting vector above the substrate at a position sufficiently distant from the dipole (See Supplemental Material [24]).

TABLE III. Wavelength-averaged Purcell factors calculated from FEM results

| Model | Purcell factor | | |
| --- | --- | --- | --- |
| | $x$-dipole | $z$-dipole | Total |
| B | 0.90 | 2.00 | 1.66 |
| C | 7.48 | 0.56 | 7.46 |

The Purcell factor of model C is much higher than that of model B. This is because the LSP mode of the nanohole supports larger EMLDOS in the hole where the emitter is placed, compared to the flat surface with propagating SPPs. In comparison with the experimental Purcell factors of corresponding structures, namely samples B and C, the simulated values show a qualitative agreement that the emitter embedded in a metal nanohole gives rise to stronger Purcell enhancement. The discrepancy of the absolute factors between the experiment and simulation can be understood from the exact shape difference, the more lossy permittivity in the actual NDs, and the lossy "gap"s instead of pure dielectrics. The shape-sensitivity can also be understood from more variations of the spectra in Samples B and C compared to Sample A. Thus, the numerical analysis by FEM revealed the relation between the Purcell factors and the enhanced EMLDOS by SPP and LSP, elucidating the experimental results.

## V. SUMMARY

The $g^{(2)}(\tau)$ measurement of individual NDs using CL photon bunching phenomenon showed that the lifetime and $g^2(0)$ of ND particles have large distributions. In the presence of a flat metal surface (Sample B), the lifetime distribution shifts to short values compared to pristine NDs (Sample A). This experimentally proves the Purcell effect due to the coupling of NV centers to SPPs. Such coupling of a dipole emitter to SPPs can be analytically formulated, which shows that the Purcell effect largely depends on the direction of the dipole and the distance from the Ag surface. These results indicate that there is a room to improve the coupling between NV centers and SPPs practically by manipulating the dipole orientation and its distance from the metal substrate. For NDs embedded in Ag nanohole (Sample C), NV centers are more strongly affected by the Purcell effect, resulting in even shorter lifetimes with a narrower distribution compared to those on flat metal surfaces (Sample B). From the evaluation using FEM simulations, it was found that the resonance peak of the in-plane LSP mode of Ag nanohole overlaps with the emission spectrum of NV. This causes a strong Purcell effect based on the coupling to the LSP confined at the Ag nanohole, where EMLDOS is strongly enhanced. Such an LSP-mediated SPP coupling geometry can open up the efficient control of the interaction between light and matter in nanoscale. The dissipated energy portion in generating SPPs can be retrieved and utilized by guiding the SPPs and converting them to photons, e.g., by using gratings [36].

The presented results and obtained insights into the coupling of NV centers to SPPs and LSPs found the basis for manipulating QEs as well as enhancing the efficiency of the light emitters such as LEDs and LDs. To maximize the efficiency of such devices, it is necessary to investigate the Purcell effect at a spatial resolution below the diffraction limit of light because the EMLDOS of the nanostructures varies in nanoscales. This measurement method, that is, the HBT interferometry for CL with both SEM and STEM operations, made it possible to investigate the Purcell effect in nanoscale as well as the dynamic information such as excitation and emission efficiencies [18]. Thus, photon bunching in CL can pave the way to unveil the bizarre light-matter interaction in far smaller dimensions than purely optical means.

**ACKNOWLEDGMENTS**

This work was financially supported by JST PRESTO (JPMJPR17P8), Murata Science Foundation, JSPS KAKENHI Grant Number 17K14321, and Research Foundation for Opto-Science and Technology. The authors thank the Suzukakedai Materials Analysis Division, Technical Department, Tokyo Institute of Technology for the structural characterization using SEM. The authors are grateful to Dr. M. Kozuma for providing us experimental equipment and Dr. T. Yuge for valuable discussion.

# Supplemental Material for

# Purcell effect of nitrogen-vacancy centers in nanodiamond coupled to propagating and localized surface plasmons revealed by photon-correlation cathodoluminescence

Sotatsu Yanagimoto[1], Naoki Yamamoto[1], Takumi Sannomiya[1, 2, *], Keiichirou Akiba[1, 3, †]

[1] Department of Materials Science and Engineering, School of Materials and Chemical Technology, Tokyo Institute of Technology, 4259 Nagatsuta, Midoriku, Yokohama, 226-8503 Japan.

[2] JST-PRESTO, 4259 Nagatsuta, Midoriku, Yokohama, 226-8503 Japan.

[3] Takasaki Advanced Radiation Research Institute, National Institutes for Quantum and Radiological Science and Technology, 1233 Watanuki, Takasaki, Gunma, 370-1292, Japan

* sannomiya.t.aa@m.titech.ac.jp

† akiba.keiichiro@qst.go.jp

**S1. Fabrication of Sample C**

Sample C was prepared through four fabrication steps, as shown in Fig. S1. To form the embedded structure, a thick Ag layer is deposited on the nanodiamond (ND) particles which are dispersed on the $SiO_2$ membrane on a NaCl substrate. This structure is transferred to a Cu transmission electron microscope (TEM) grid in an upside-down manner so that the measurement can be performed from the ND side. A detailed surface image of Sample C was taken by scanning electron microscope (SEM) (S-5500, HITACHI, Japan) to investigate the detailed surface structure. The SEM secondary electron image (Fig. S2) shows clear edge contrasts around the embedded ND and a darker contrast in the ND position, confirming that Sample C has the intended structure in which ND is embedded in the Ag substrate with its surface exposed. A similar structure with the embedded NDs fabricated through electron-beam lithography has been reported [1], but the presented method here can more easily realize the hybrid NDs-Ag nano-hole system.

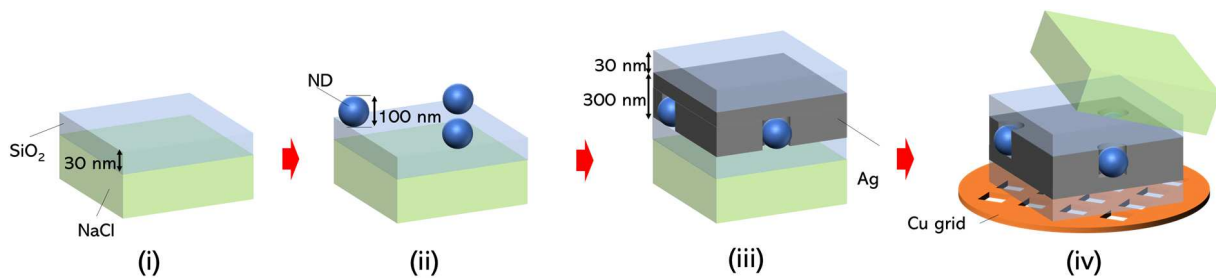

FIG. S1. Fabrication method of Sample C. (i) A 30 nm $SiO_2$ thin membrane is sputter-deposited on a cleaved surface of a NaCl crystal. (ii) NDs are dispersed on the substrate

from the solution. (iii) A 300 nm thick Ag film and a 30 nm thick $SiO_2$ membrane are sputter-deposited on the sample to embed the NDs in Ag. The $SiO_2$ membranes are to protect the Ag surface from oxidation in water used in the next process. (iv) To remove NaCl and transfer the membrane to a Cu grid, the sample is put upside down on the Cu grid in water.

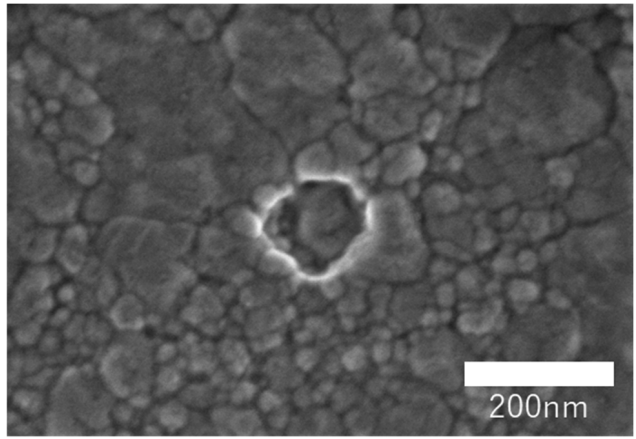

FIG. S2. SEM secondary electron image of Sample C taken at an accelerating voltage of 1.0 kV. It can be confirmed that the ND is embedded in the Ag nanohole with a diameter of about 130 nm.

## S2. Derivation of second-order correlation function $g^{(2)}(\tau)$ using two-level system

When a single incident electron excites multiple emission centers at the same time, a large photon bunching phenomenon occurs in the emitted light. Using a simple two-level system as a model of the emission center, we derive the expression of the second-order correlation function $g^{(2)}(\tau)$ that represents bunching. As shown in Fig. S3(a), it is assumed that light emission is generated by the electric dipole transition between the ground state (1) and the excited state (2). $G$ is the excitation rate that causes a $1 \rightarrow 2$ transition by the incident electron, and $\gamma_r$, $\gamma_{nr}$ and $\gamma_p$ are the transition rates of the $2 \rightarrow 1$ transition associated with the radiative process, non-radiative process, and plasmons, respectively. Now we consider a situation where the electron beam current is sufficiently weak such that an excitation process of one incident electron has no influence on the next electron excitation and that a linear sum of each process can describe the whole phenomenon. For the electron beam current $I_b$, the average time interval $t_b$ between consecutive electron incidents is given by $t_b = (I_b/e)^{-1}$. If the excitation probability of one incident electron is $\eta$, the average time interval $t_{ex}$ at which $1 \rightarrow 2$ excitation occurs is $(t_b/\eta)$. Let $G(t)$ be the function that expresses such excitation and can be written as

$$G(t) = \sum_{i=1}^{N} N_i^{ex} \delta(t - t_i) \qquad (S1)$$

where $N_i^{ex}$ is the number of centers that are simultaneously excited at time $t_i$. Using the average excitation number of emission centers per electron $N_{ex}$, the average excitation rate $G$ is given by

$$G = \frac{N_{ex}}{t_{ex}} = \eta \frac{N_{ex}}{t_b} = \eta \left( \frac{I_b}{e} \right) N_{ex} \qquad (S2)$$

If the number of centers staying at the excited state (2) at time $t$ is $n(t)$, the rate equation is expressed as

$$\frac{dn(t)}{dt} = G(t) - (\gamma_r + \gamma_{nr} + \gamma_p)n(t) = G(t) - \gamma n(t) \quad (S3)$$

where $\gamma = \gamma_r + \gamma_{nr} + \gamma_p$ is the total decay rate considering radiative, non-radiative and plasmon decay paths.

Between the time interval of $t_i$ and $t_{i+1}$, no excitation occurs, thus the rate equation from a single excitation event can be described as

$$\frac{dn_i(t)}{dt} = -\gamma n(t) \quad (S4)$$

and the solution of the single process with the initial condition that $n_i(0) = N_i^{ex}$ at $t = t_i = 0$, is given by

$$n_i(t) = N_i^{ex} e^{-\gamma t} \quad (S5)$$

This situation is schematically illustrated in Fig. S3(b). The single excitation-emission process is repeated for each randomly incident electron. When $N$ excitation events occur over a sufficiently long time range, the total number $n(t)$ is expressed by

$$n(t) = \sum_{i=1}^{N} N_i^{ex} e^{-\gamma(t-t_i)} \theta(t - t_i) \quad (S6)$$

where $\theta(t - t_i)$ is a step function, i.e., $\theta(t - t_i) = 1$ when $t \geq t_i$ and $0$ when $t < t_i$. The emission intensity of cathodoluminescence (CL) is proportional to the number of photons emitted from the centers per unit time and thus expressed as

$$I(t) = \gamma_r n(t) \quad (S7)$$

where the photodetection efficiency of the CL system is ignored.

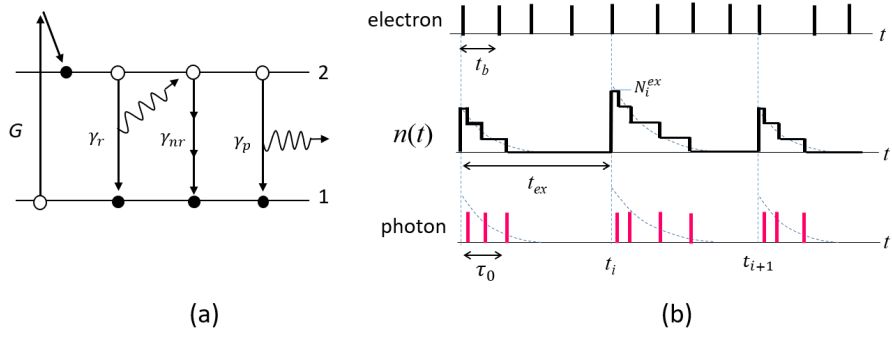

FIG. S3 (a) Two-level system as a model of an emission center. (b) Classical picture of the relation among electron incidence, number of excited emission centers $n(t)$ and photon emission.

The second order correlation function $g^{(2)}(\tau)$ in the Hanbury Brown-Twiss interferometry is defined by

Eq. (2) in the main manuscript. Using Eq. (S7), this is rewritten as

$$g^{(2)}(\tau) = \frac{\langle I(t)I(t+\tau)\rangle}{\langle I(t)\rangle\langle I(t+\tau)\rangle} = \frac{\langle n(t)n(t+\tau)\rangle}{\langle n(t)\rangle \langle n(t+\tau)\rangle} \qquad (S8)$$

According to the ergodic theorems, $\langle n(t)\rangle$ can be calculated as the time average of $n(t)$ :

$$\langle n(t)\rangle = \lim_{T\to\infty}\frac{1}{T}\int_0^T \sum_{i=1}^N N_i^{ex} e^{-\gamma(t-t_i)}\,\theta(t-t_i)dt = \lim_{T\to\infty}\frac{1}{T}\sum_{i=1}^N N_i^{ex}\,\frac{1}{\gamma} = \frac{N_{ex}}{\gamma t_{ex}} \qquad (S9)$$

where $N_{ex} = \frac{1}{N}\sum_{i=1}^N N_i^{ex}$ and $\lim_{T\to\infty}\frac{N}{T} = \frac{1}{t_{ex}}$ . It is evident that $\langle n(t)\rangle = \langle n(t+\tau)\rangle$ . Furthermore, using

Wiener-Khinchin theorem, the correlation function $\langle n(t)n(t+\tau)\rangle$ is calculated as follows,

$$\langle n(t)n(t+\tau)\rangle = \lim_{T\to\infty}\frac{1}{T}\int_0^T n(t)n(t+\tau)dt = \lim_{T\to\infty}\frac{1}{2\pi}\int_{-\infty}^\infty \frac{1}{T}|n_\omega(\omega)|^2 e^{i\omega\tau}d\omega \qquad (S10)$$

Here $n_\omega(\omega)$ is the Fourier transform of $n(t)$ and expressed as

$$n_\omega(\omega) = \int_{-\infty}^{\infty} n(t)e^{-i\omega t}dt = \frac{1}{\gamma + i\omega}\sum_{k=1}^{N} N_k^{ex}\, e^{i\omega t_k} \tag{S11}$$

Inserting (S11) to (S10), and after some algebra, we get

$$\langle n(t)n(t+\tau)\rangle = \lim_{T\to\infty}\frac{1}{2\pi T}\int_{-\infty}^{\infty}\frac{1}{\gamma^2+\omega^2}\cdot\left[\sum_{k=j}^{N}(N_k^{ex})^2 + \sum_{k\neq j}^{N}N_k^{ex}N_j^{ex}e^{i\omega(t_k-t_j)}\right]e^{i\omega\tau}d\omega$$

$$= \frac{1}{2\gamma t_{ex}}e^{-\gamma|\tau|}\lim_{N\to\infty}\frac{1}{N}\sum_{k=1}^{N}(N_k^{ex})^2 + \frac{N_{ex}^2}{\gamma^2\, t_{ex}^2} \tag{S12}$$

Finally, $g^{(2)}(\tau)$ is derived as follows,

$$g^{(2)}(\tau) = \frac{\gamma t_{ex}}{2}\xi e^{-\gamma|\tau|} + 1 \tag{S13}$$

where $\xi$ is defined by

$$\xi = \left(\lim_{N\to\infty}\frac{1}{N}\sum_{k=1}^{N}(N_k^{ex})^2\right)\bigg/\left(\lim_{N\to\infty}\frac{1}{N}\sum_{k=1}^{N}N_k^{ex}\right)^2 = \left(\lim_{T\to\infty}\frac{1}{N}\sum_{k=1}^{N}(N_k^{ex})^2\right)\bigg/N_{ex}^2 \tag{S14}$$

where $\xi = 1$ if $N_k^{ex} = N_{ex}$ for all $k$. Fluctuations in $N_k^{ex}$ around $N_{ex}$ causes an increase of $\xi$, such that

$\xi \geq 1$. Using Eq.(S2) and $\gamma = 1/\tau_0$, we get

$$g^{(2)}(0) - 1 = \frac{\gamma t_{ex}}{2}\xi = \frac{e\xi}{2\eta\tau_0 I_b} \tag{S15}$$

And the cathodoluminescence intensity is given by

$$\langle I(t)\rangle = \gamma_r\langle n(t)\rangle = \frac{\gamma_r}{\gamma}G = \frac{\gamma_r}{\gamma}\eta N_{ex}\left(\frac{I_b}{e}\right) \tag{S16}$$

S7

**S3. Distribution of lifetime**

Here we describe how the distribution of the lifetime $\tau_0$ is modified with an additional decay rate, e.g. $\gamma_p$, by the presence of the metallic structure. We first define the distribution of the decay rate $\gamma$ which is now expressed as a probability function $p(\gamma)$. Since the distribution of $\gamma$ is related to the distribution of the size and shape of NDs [2,3], we can reasonably assume that the function $p(\gamma)$ has a gaussian distribution. Thus the $p(\gamma)$ with the center peak at $\gamma_n$ is expressed as $p(\gamma, \gamma_n) \propto \exp[-(\gamma - \gamma_n)^2/2\sigma^2]$, where $\sigma$ is the standard deviation. This distribution can be converted in the scale of by $\tau = 1/\gamma$. In Fig. S4(a), three gaussian distributions in the decay rate with the center peak shifts, namely $\gamma_1$, $\gamma_2$, $\gamma_3$, are shown. The corresponding axis-converted distributions are plotted in panel (b). These graphs show that an addition of a constant value in $\gamma$, or shift of the peak, results in a narrower distribution in the scale of $\tau$ (panel (b)). Such distributions nicely reproduce the histograms of Fig. 5 in the main manuscript, evidencing the addition of the decay rate by the presence of the metallic structure.

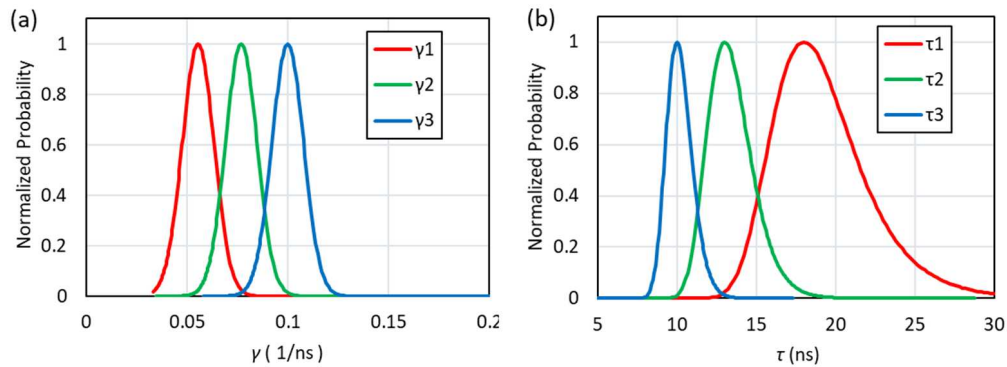

FIG. S4. (a) Three different decay rate distributions with the same distribution width and with the center positions shifted by certain values. (b) Corresponding lifetime distributions.

## S4. Relation between $g^{(2)}(\tau)$ and beam current

Using Samples A and B, $g^{(2)}(\tau)$ was measured while changing the incident beam current on the same ND particle. Figures S5(a-c) and (d-f) correspond to Sample A and B, respectively. We evaluated the dependence of $g^{(2)}(0)$ and lifetime on the beam current. The $g^{(2)}(0)$ values are extracted by fitting the $g^{(2)}(\tau)$ curves of Fig. S5(a) and (d). It can be confirmed that $\left(g^{(2)}(0) - 1\right)$ is inversely proportional to the current density, as shown in panels (b) and (e), which is in agreement with Eq. (S15) and the previous reports [4]. Fitting the experimental results with Eq. (S15), as shown in Fig. S5(b) and (e), provides the excitation probability per incident electron, which was calculated as $\eta_A \sim 30\%$ and $\eta_B \sim 40\%$ using $\xi = 1.1$. These values are reasonable considering the 100 nm thickness of the diamond and the acceleration voltage of 80 kV [5]. In contrast to $g^{(2)}(0)$, no apparent dependence of the lifetime on the beam current was observed (panels (c) and (f)). From this, we can confirm that the beam current, especially in the range of 17 to 60 pA, does not affect the result of the lifetime measurement.

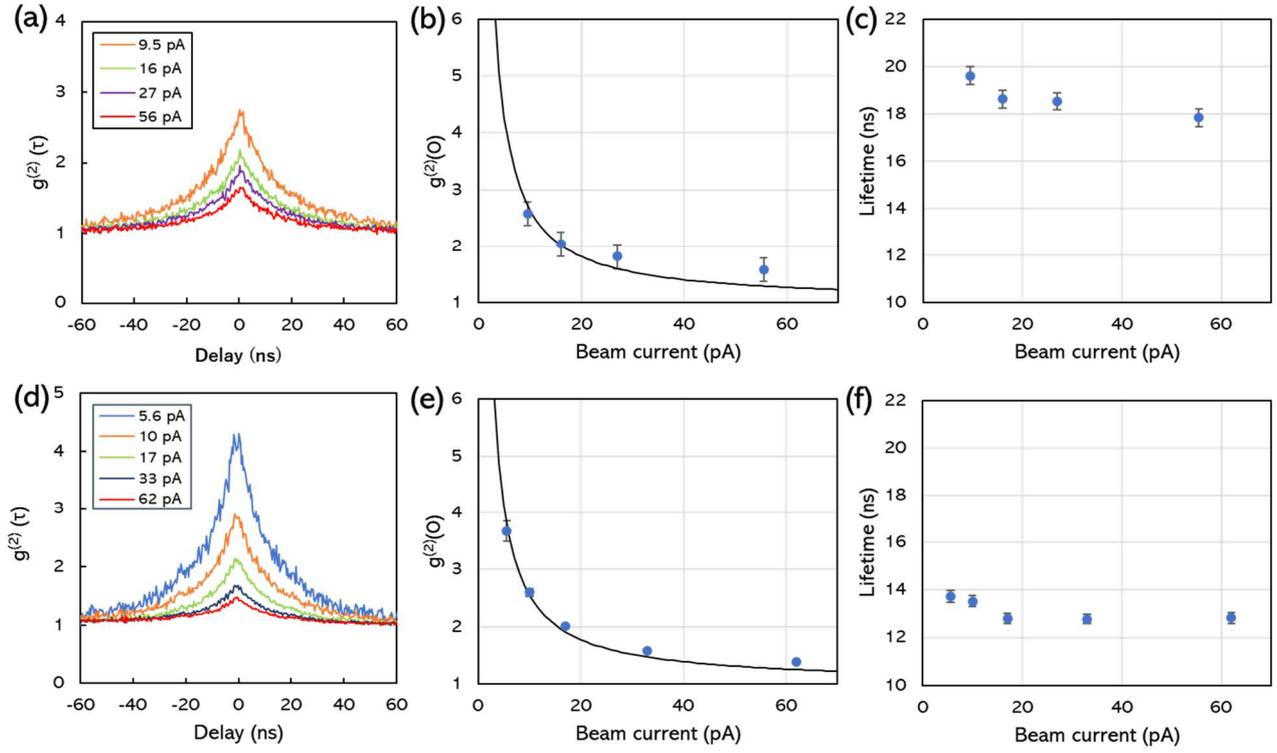

FIG. S5. Beam current dependence of $g^{(2)}(\tau)$ curve. (a-c) and (d-f) are the results of Sample A and B, respectively. (a,d) $g^{(2)}(\tau)$ curves measured by changing the beam current. (b,c,e,f) Beam current dependence of (b,e) $g^{(2)}(0)$ and (c,f) lifetime, respectively. In panel (b,e), the fitted curve using Eq. (S15) is shown as the black line.

## S5. Size effect of the ND sphere

The ND particles used in the experiment have an average diameter of 100 nm and distribution of 70 to 125 nm [900174-5ML, Sigma-Aldrich, America]. We investigated the size effect on the Purcell factors by electromagnetic field calculation using FEM. An electric dipole was placed in the center of a spherical ND particle with a 2.5 nm graphite layer on the surface, and the simulations were performed for diameters of 70 to 130 nm. The wavelength-integrated Purcell factor was obtained by using Eq. (4) and normalized with the result of the diameter of 100 nm. Here, to observe purely the size effect, models with only the ND particles were used. Figure S6 shows the normalized Purcell factors as a function of the particle diameter. The simulation demonstrates that the Purcell factor can vary by the particle size. In the experiment, additional effects related to the shape of the particle and the location of the dipole would be included. Due to these combined effects, even bare NDs have a wide lifetime distribution.

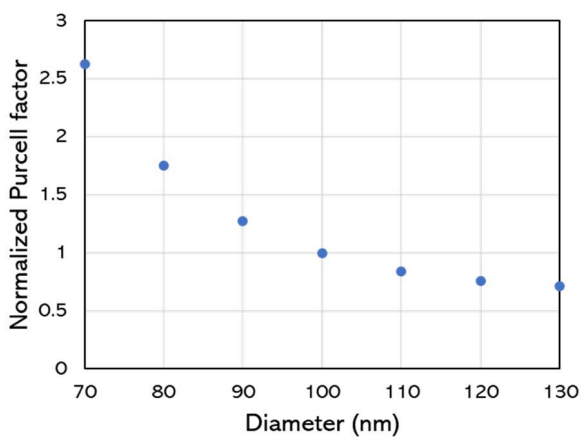

FIG. S6. Size dependence of the Purcell factor with the ND particle diameter ranging from 70 to 130 nm. The Purcell factors are normalized by the result of 100 nm particles.

## S6 .CL spectrum of individual NDs

It is expected that the spectral shapes of ND emission in Sample B and C are modified from the bare NDs due to the coupling to the plasmon modes. We here compare the spectra of NDs of Samples A, B and C. We performed the measurement with a fixed beam current at 60pA. The exposure time to measure the $g^{(2)}(\tau)$ curve and spectrum is 100 seconds. Figure. S6 shows the SEM or STEM bright field images (leftmost panel), CL-maps (second panel from the left), $g^{(2)}(\tau)$ curves (second panel from the right), and CL spectra (rightmost panel) for representative ND particles in the same field of view for each sample. The selected ND particles are indicated by circles, of which color corresponds to the $g^{(2)}(\tau)$ curves and the spectra.

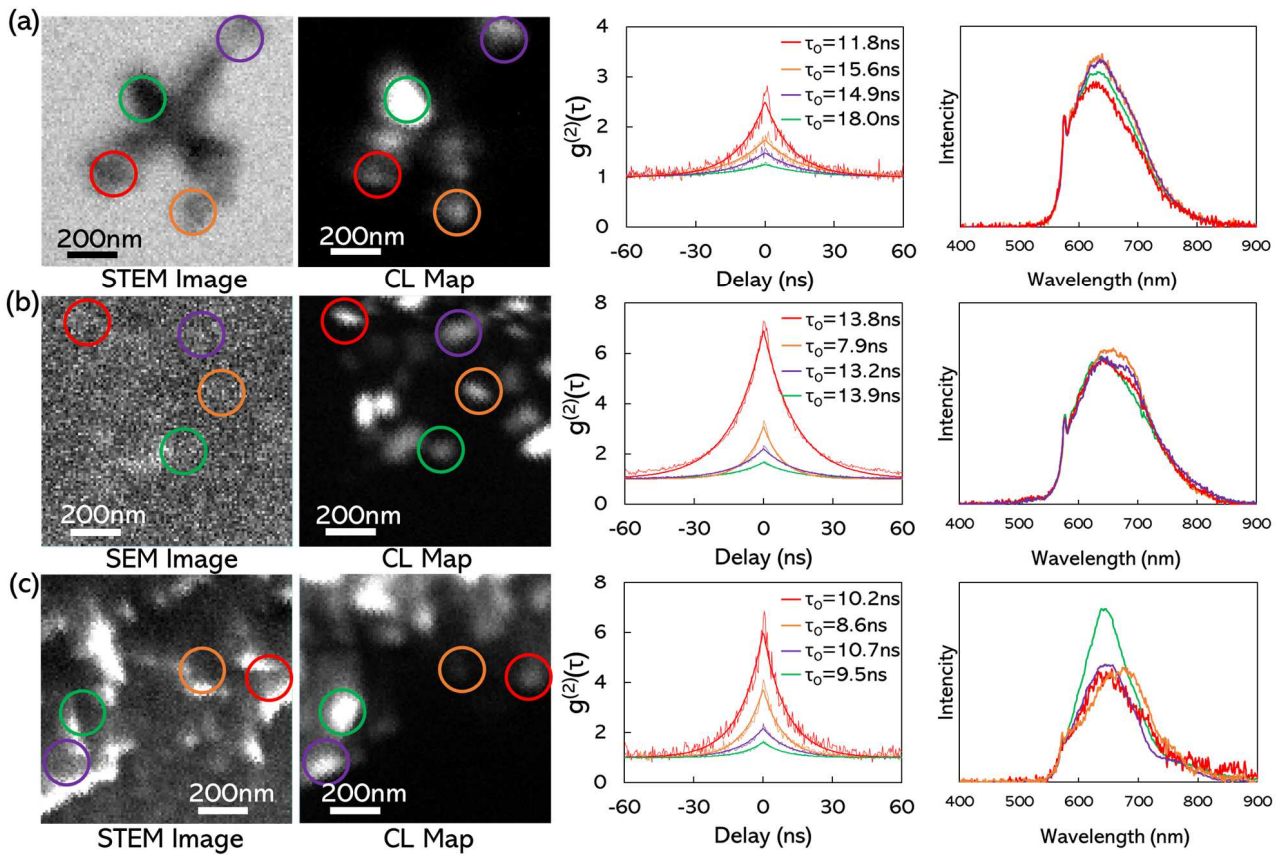

FIG. S7 .CL-HBT measurement results of Samples (a) A, (b) B, and (c) C, respectively. From left to right: SEM or STEM image, CL panchromatic image, $g^{(2)}(\tau)$ plots with fitted curves, and CL spectra of individual NDs. The color of the $g^{(2)}(\tau)$ curve and the spectrum of each sample corresponds to the color of the circle in the map which indicates

the measured NDs. To compare the shape of the spectrum, we normalized the intensity at the zero-phonon line (ZPL) with the wavelength of 575 nm.

The measurement results show that the shape of the spectrum depends on the presence of the Ag structure.

While the spectral shapes of all the "bare" particles in Sample A (NDs on $SiO_2$ membrane) are almost identical,

more variations of the spectra have been observed in the samples with Ag structures (Sample B and C). The

variation is relatively smaller for Sample B then Sample C because the density of states of the SPP on a flat

Ag surface is a monotonic function of the wavelength in this optical region. On the other hand, the variation is more distinct for Sample C because of the strong coupling to and the resonance of the localized surface plasmon modes.

## S7. Relation between $\tau_0$ and $g^{(2)}(0)$

A scatter plot of the lifetime and $g^{(2)}(0)$ of Samples A, B and C are shown in Fig. S6. Here, only the measurement results with the beam current of 17 pA are used because $g^{(2)}(0)$ value depends on the current density. As shown in Fig. S8, the $g^{(2)}(0)$ tends to be distributed at higher values for Sample B with the Ag

substrate and even higher for Sample C with the embedded structure, compared to the pristine NDs of Sample

A. This tendency is consistent with the formula in Eq. (S15). For a given beam current, a shorter lifetime means less overlaps of the photons of subsequent emission events, thus increasing the $g^{(2)}(0)$ value. There is also a possible influence of secondary electrons on Samples B and C with the metal structures.

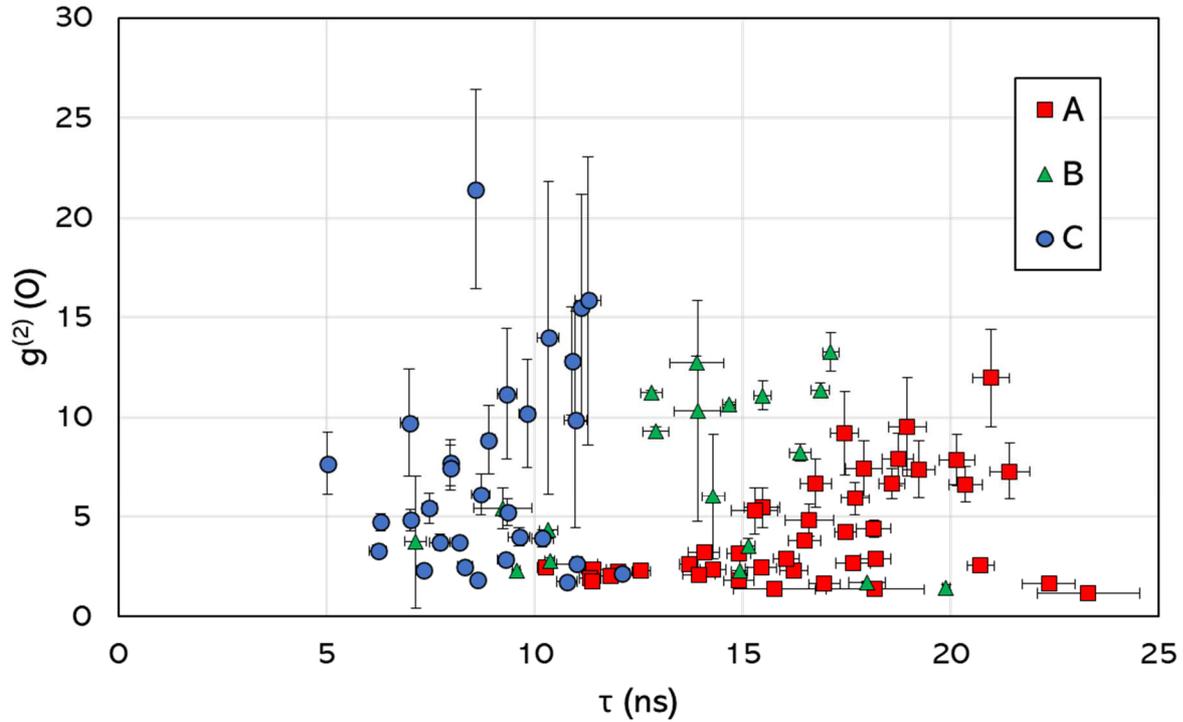

FIG. S8. Relation between $g^{(2)}(0)$ and lifetime. The number of plots for Samples A, B, and C is 42, 19, and 30, respectively. Error estimations due to the fitting are shown as error bars.

## S8. Amorphous/graphitized layer around nanodiamond

We observed the amorphous or graphitized layer surrounding the ND particle using JEM-2100F and R005 TEM instruments with field emission guns operated at acceleration voltages of 80 kV and 200kV, respectively.

The sample was prepared by dispersing NDs on a microgrid. The amorphous/graphitized layer of 2-3 nm is confirmed in the TEM images as shown in Fig. S9, and the thickness of the layer is in agreement with the previous report [6].

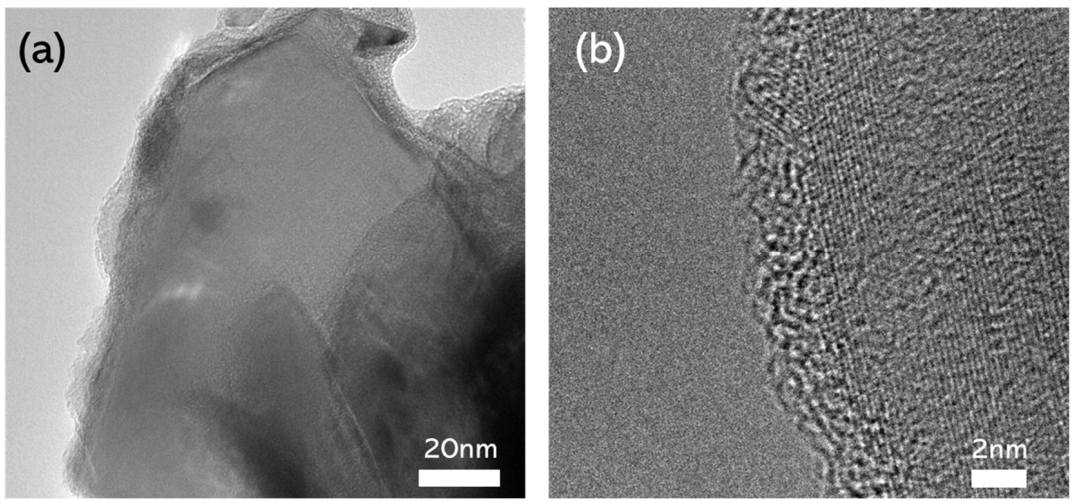

FIG. S9. (a) and (b) are TEM images of NDs. An amorphous or graphitized layer of 2-3 nm around the diamond crystal is observed in both images.

## S9. Analytical evaluation of dipoles on a substrate

Figure S10 shows the spectral plots of the analytically calculated Purcell factors of dipoles placed at $z_0 = 50$ nm from a silver surface. The ratio of the radiative and total Purcell factors changes depending on the wavelength. Since the detection is only the radiative portion of the energy, we weighted the Purcell factor for each radiative energy when integrating the Purcell factor over the wavelength in Eq. (4) in the main manuscript. The radiative portion of the energy is proportional to the radiative Purcell factor (blue line in Fig. S10) [7].

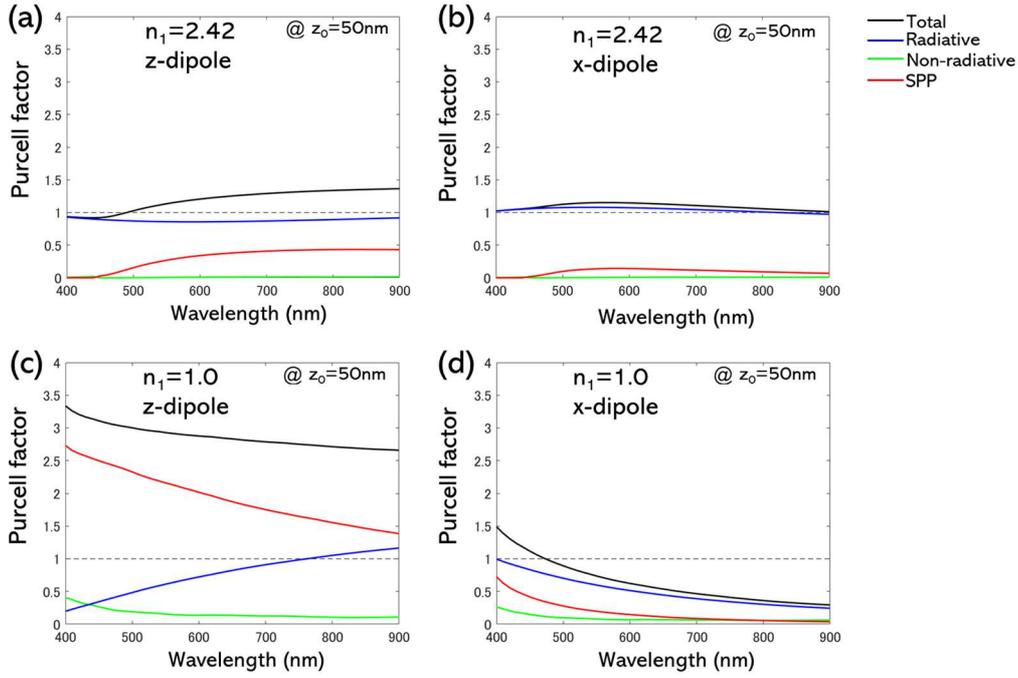

FIG. S10. Spectral plots of Purcell factors calculated by analytical formula. (a,b) Purcell factors of the $z$- and $x$-dipoles in diamond ($n_1$=2.42). (c,d) Purcell factors of the $z$- and $x$-dipoles in vacuum ($n_1$=1.0).

## S10. Details of FEM models and validation of the evaluation method

In the finite element method (FEM) calculations, electric point dipoles with zero volume were used as the excitation source. The electric field at the dipole position is evaluated in the calculation of the Purcell factor [8,9]. Since the calculation space is discretized by finite-size volumes in FEM, the volume-integral in Eq. (8) was performed on a spherical shell in a cylindrical shell with a radius of 4 nm and a shell thickness of 2 nm surrounding the point dipole. This is based on the assumption that the electric field intensity at the position of the dipole is linear to the field in its vicinity. The FEM Model B is similar to the analytical models described by Eq. (3). However, the main difference is that the dipole in the FEM model is surrounded by a dielectric medium of a spherical particle.

Here, we calculate the exact same model as the analytical one, namely a simple dipole on a flat substrate without the presence of the dielectric particle, to validate our FEM evaluation by comparing the results with the analytical one. The spectra of the Purcell factors at $z_0 = 50$ nm obtained by FEM are shown in Fig. S11(b), which nicely reproduces the analytical calculation of Fig. S11(a).

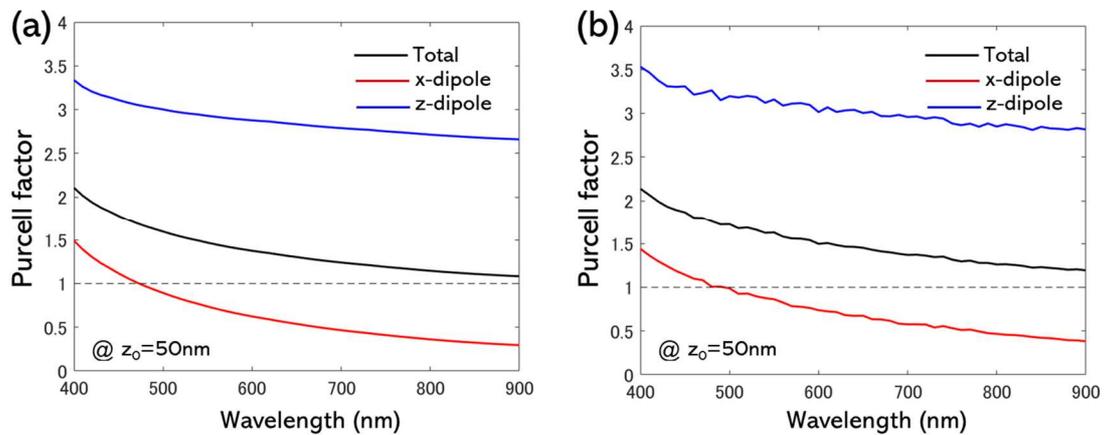

FIG. S11. Comparison between the analytical and FEM models for a dipole on a flat substrate in vacuum without dielectric particle. (a) Spectra of the Purcell factors obtained from the analytical formula, which correspond to those of black lines in FIG. S10 (c) and (d). (b) Spectra of the Purcell factors obtained from the FEM simulation.

## S11. FEM

To investigate the change in the Purcell factor depending on the position and direction of the dipoles in NDs, FEM simulation was performed for different dipole orientations and locations as shown in Fig. S12. In these models, the dipoles are placed either at the center of the particle or displaced by 25 nm from the center along the axes. We define the origin of the coordinate at the particle center and the dipoles are located at seven different positions: $(x_0, y_0, z_0)$ = (0,0,0), (±25 nm, 0, 0), (0, ±25 nm, 0), (0, 0, ±25 nm). At each position, three dipole orientations are set along the $x$-, $y$-, and $z$-axes, which results in 21 different models. By considering the symmetry of the structure, this calculation can be reduced to nine models. We therefore performed the calculations for the dipoles located at the top (0, 0, +25 nm), center (0,0,0), side (0, +25 nm, 0), and bottom (0, 0, -25 nm) of the ND.

The spectra of the Purcell factors obtained by simulation are shown in Fig. S13. For Model B, the closer the dipole is to the substrate, the larger the Purcell factor of the $z$-dipole. For Model C, the Purcell factor changed significantly depending on the position of the dipole, especially for the $z$-dipole (blue lines). As for the $z$-dipole in the ND (blue curve) in Model C, the Purcell factor at the wavelength of 430 nm is five times higher for the $z$-dipole at the top than at the bottom.

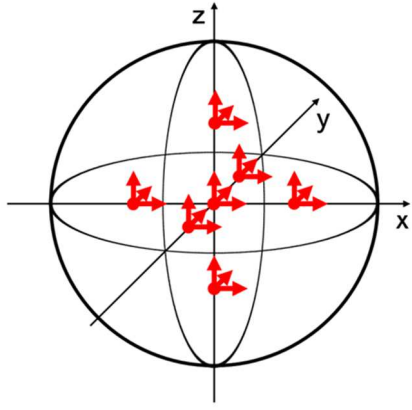

FIG. S12. Schematic diagram of the positions of 21 dipoles considered in FEM simulation.

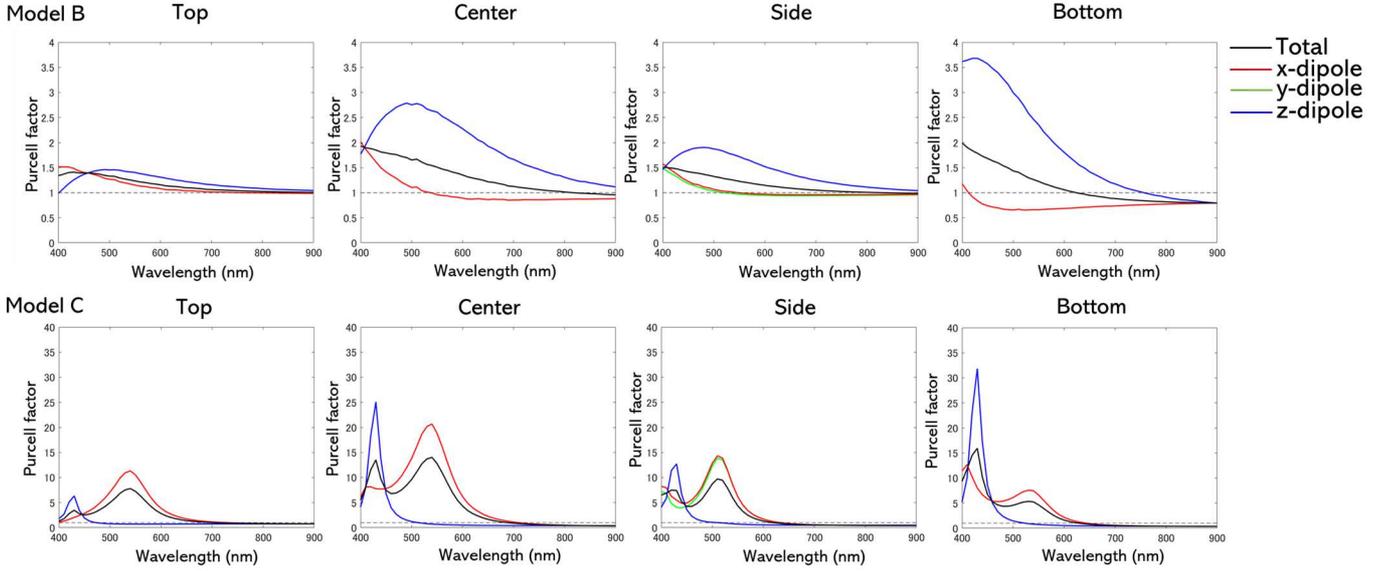

FIG. S13. Spectra of Purcell factors at various positions and directions of the dipole within the particle. The positions are defined as : Top $(x_0, y_0, z_0) = (0,0,+25[\text{nm}])$, Center $(x_0, y_0, z_0) = (0,0,0)$, Side $(x_0, y_0, z_0) = (+25[\text{nm}],0,0)$, and Bottom $(x_0, y_0, z_0) = (0,0,-25[\text{nm}])$.